# DDPM based X-ray Image Synthesizer

Thulana Abeywardane
C0861333
*Lambton College*
Toronto, CA
c0861333@mylambton.ca

Tomson George
C0857730
*Lambton College*
Toronto, CA
c0857730@mylambton.ca

Praveen Mahaulpatha
C0860583
*Lambton College*
Toronto, CA
c0860583@mylambton.ca

*Abstract* — Access to high-quality datasets in the medical industry limits machine learning model performance. To address this issue, we propose a Denoising Diffusion Probabilistic Model (DDPM) combined with a UNet architecture for X-ray image synthesis. Focused on pneumonia medical condition, our methodology employs over 3000 pneumonia X-ray images obtained from Kaggle for training. Results demonstrate the effectiveness of our approach, as the model successfully generated realistic images with low Mean Squared Error (MSE). The synthesized images showed distinct differences from non-pneumonia images, highlighting the model's ability to capture key features of positive cases. Beyond pneumonia, the applications of this synthesizer extend to various medical conditions, provided an ample dataset is available. The capability to produce high-quality images can potentially enhance machine learning models' performance, aiding in more accurate and efficient medical diagnoses. This innovative DDPM-based X-ray photo synthesizer presents a promising avenue for addressing the scarcity of positive medical image datasets, paving the way for improved medical image analysis and diagnosis in the healthcare industry.

## I. Introduction

Medical imaging plays a critical role in modern healthcare, aiding in the diagnosis and treatment of various medical conditions. Among the different imaging modalities, X-ray analysis is widely used due to its cost-effectiveness and non-invasive nature. In the context of respiratory diseases, pneumonia stands out as a common and life-threatening lung infection. With the emergence of COVID-19 in late 2019, the world witnessed a global pandemic, further highlighting the significance of accurate and efficient chest X-ray analysis for detecting lung infections.

The successful application of machine learning models for medical image analysis heavily relies on the availability of high-quality and diverse datasets for training. However, obtaining an adequate number of positive cases, such as pneumonia-infected X-ray images, remains a persistent challenge in the medical industry. Several factors contribute to this scarcity. Firstly, acquiring real patient data for research purposes can be infeasible due to constraints on time, labor, or financial resources. Secondly, privacy concerns and strict data regulatory legislations limit the sharing of medical imaging data between different institutions. As a result, training machine learning models for pneumonia detection becomes increasingly difficult, as the models lack access to sufficient positive cases.

In light of these challenges, our project aims to address the scarcity of positive pneumonia cases for training by developing a Deep Density Priors Model (DDPM) based X-ray photo synthesizer. The primary objective is to generate realistic and diverse X-ray images of pneumonia, thus supplementing the limited positive cases in existing datasets.

Pneumonia is classified as a fatal lower respiratory infection under the acute diseases category and has been reported as a leading cause of deaths globally. For instance, in 2017, it accounted for approximately 15% of child deaths during the year. Additionally, older individuals are particularly vulnerable to pneumonia, which can lead to critical conditions. However, early diagnosis and appropriate treatment can significantly reduce associated risks. [3]

Radiologists typically analyze chest X-ray images to identify infiltrates or white spots in the lungs, indicative of infections like pneumonia or COVID-19. This analysis also helps identify complications, such as pleural effusions, which involve an excess of fluid surrounding the lungs. Given the complexities of identifying subtle abnormalities in chest X-rays, the interpretation process requires expert knowledge and experience, making it a challenging task.

To aid medical professionals in this diagnostic process, computer-aided solutions have been explored. The use of machine learning algorithms can serve as supportive tools, helping to reduce human error and improve the efficiency of the diagnosis process.

The scope of this project is focused on pneumonia detection, given its prevalence and impact on global health. By creating a DDPM-based X-ray photo synthesizer, we aim to generate synthetic X-ray images that closely resemble real pneumonia cases. These generated images will serve as valuable additions to medical image datasets, effectively augmenting the limited pool of positive cases available for training machine learning models.

The significance of this project lies in its potential to enhance the performance and accuracy of pneumonia detection models. With access to a more extensive and diverse set of positive cases, machine learning algorithms can be better equipped to identify pneumonia infections accurately, leading to improved patient outcomes and more efficient medical diagnoses.

To achieve our objectives, we will train the DDPM model on a dataset of over 3000 pneumonia X-ray images obtained from Kaggle, ensuring that the model learns the intricate features and patterns characteristic of pneumonia infections.

In the subsequent sections, we present the methodology used in detail, the experimental results demonstrating the effectiveness of our approach, and conclude by discussing the project's accomplishments and potential future improvements.

## II. METHODOLOGY

The methodology employed in this project aims to create a state-of-the-art X-ray photo synthesizer using diffusion models, with a specific focus on the Denoising Diffusion Probabilistic Model (DDPM). By leveraging these advanced probabilistic models, we seek to generate highly realistic and diverse X-ray images that closely resemble positive pneumonia cases. The fundamental principle behind diffusion models is to iteratively transform a noise-initialized image into a target image, allowing for the generation of complex and coherent patterns.

Diffusion models, as a class of generative models, have shown great promise in image synthesis tasks. They are particularly well-suited for medical imaging applications, where the scarcity of positive cases poses significant challenges in training machine learning models effectively. The application of diffusion models enables us to create synthetic images that closely mimic real medical cases, effectively augmenting limited datasets and enhancing the performance of diagnostic algorithms.

At the core of our methodology lies the Denoising Diffusion Probabilistic Model (DDPM). DDPM combines denoising autoencoders with probabilistic modeling to produce high-quality images with remarkable fidelity. By adopting this cutting-edge approach, our X-ray photo synthesizer can generate diverse and accurate images of pneumonia, effectively mitigating the data scarcity issue commonly faced in medical imaging research.

In the subsequent sections, we will delve into the intricacies of diffusion models and provide an in-depth explanation of the Denoising Diffusion Probabilistic Model (DDPM). Additionally, we will detail the dataset used for training, the training process, and the evaluation metrics employed to assess the performance of our DDPM-based X-ray photo synthesizer. Through this methodology, we aim to contribute to the advancement of medical image synthesis and provide a valuable tool for improved diagnosis and patient care in the field of healthcare.

### A. DENOISING DIFFUSION PROBABILISTIC MODEL

A diffusion probabilistic model is a parameterized Markov chain trained using variational inference to produce samples matching the data after a finite time. Transitions of this chain are learned to reverse a diffusion process, which is a Markov chain that gradually adds noise to the data in the opposite direction of sampling until the signal is destroyed. When the diffusion consists of small amounts of Gaussian noise, it is sufficient to set the sampling chain transitions to conditional Gaussians too, allowing for a particularly simple neural network parameterization. Also, for the hyper parameters, one must choose the variances $\beta_t$ of the forward process and the model architecture and Gaussian distribution parameterization of the reverse process. [2]

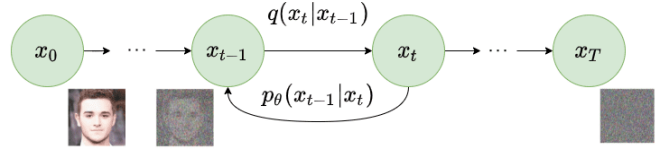

Figure 1 Forward diffusion process. Image modified by Ho et al. 2020

**Forward Process:** DDPM defines the forward diffusion process as a Markov Chain where Gaussian noise is added in successive steps to obtain a set of noisy samples. Consider $q(x_0)$ as the uncorrupted (original) data distribution. Given a data sample, $x_0 \sim q(x_0)$, a forward noising process p, which produces latent x1 through $x_T$ by adding Gaussian noise at time t, is defined as follows:

$$q(x_t \mid x_{t-1}) = \mathcal{N}\left(x_t; \sqrt{1-\beta_t} \cdot x_{t-1}, \beta_t \cdot \mathbf{I}\right), \forall t \in \{1, \ldots, T\},$$
(1)

where T and $\beta_1, \ldots, \beta_T \in [0, 1)$ represent the number of diffusion steps and the variance schedule across diffusion steps, respectively. I is the identity matrix, and N (x; μ, σ) represents the normal distribution of mean μ and covariance σ. Considering $\alpha_t = 1 - \beta_t$ and $\bar{\alpha}_t = \prod_{s=0}^{t} \alpha_s$, one can directly sample an arbitrary step of the noised latent conditioned on the input x0 as follows:

$$q(\mathbf{x}_t \mid \mathbf{x}_0) = N\left(\mathbf{x}_t; \sqrt{\bar{\alpha}_t}\mathbf{x}_0, (1-\bar{\alpha}_t)\mathbf{I}\right)$$
(2)

$$\mathbf{x}_t = \sqrt{\bar{\alpha}_t}\mathbf{x}_0 + \sqrt{1-\bar{\alpha}_t}\epsilon.$$
(3)

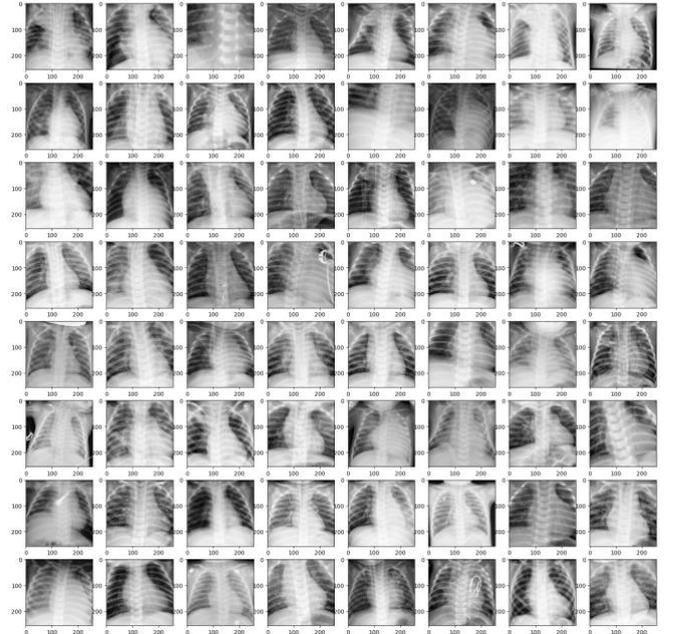

Figure 2: Original Images

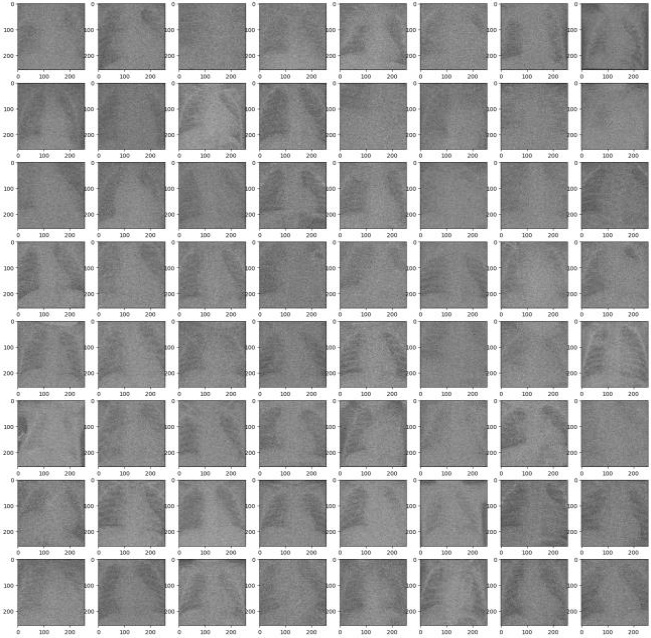

Figure 3: Adding 25%

Figure 2 and figure 3 represent the images at time step 0 and after 25% of noise adding. Looking at this, we can see the noise addition is happening systematically from a gaussian distribution and not completely random. This allows the network to predict the noise more easily than from a completely random distribution, where it has to learn the mean and variance too.

**Reverse Process:** Leveraging the above definitions, we can approximate a reverse process to get a sample from q (x0). To this end, we can parameterize this reverse process by starting at p (xT ) = N (xT ; 0, I) as follows:

$$p_\theta(\mathbf{x}_{0:T}) = p(\mathbf{x}_T) \prod_{t=1}^{T} p_\theta(\mathbf{x}_{t-1} | \mathbf{x}_t)$$

(4)

$$p_\theta(\mathbf{x}_{t-1} | \mathbf{x}_t) = \mathcal{N}(\mathbf{x}_{t-1}; \mu_\theta(\mathbf{x}_t, t), \Sigma_\theta(\mathbf{x}_t, t)).$$

(5)

To train this model such that p (x0) learns the true data distribution q (x0), we can optimize the following variational bound
on negative log-likelihood:

$$\mathbb{E}[-\log p_\theta(\mathbf{x}_0)] \leq \mathbb{B}_q \left[-\log \frac{p_\theta(\mathbf{x}_{0:T})}{q(\mathbf{x}_{1:T} | \mathbf{x}_0)}\right]$$
$$= \mathbb{E}_q \left[-\log p(\mathbf{x}_T) - \sum_{t \geq 1} \log \frac{p_\theta(\mathbf{x}_{t-1} | \mathbf{x}_t)}{q(\mathbf{x}_t | \mathbf{x}_{t-1})}\right]$$
$$= -L_{VLB}.$$

(6)

Diffusion Models in Medical Imaging: A Comprehensive Survey Ho et al. [2] found it better not to directly parameterize μθ (xt, t) as a neural network but instead to train a model εθ (xt, t) to predict ε. Hence, by reparametrizing Equation (6), they proposed a simplified objective as follows:

$$L_{simple} = E_{t, x_0, \epsilon}\left[\|\epsilon - \epsilon_\theta(x_t, t)\|^2\right],$$

(7)

where the authors draw a connection between the loss in Equation (6) to generative score networks in Song et al. [5].

### B. UNET ARCHITECTURE

The U-Net takes just an image as input and produces another image as output. The neural network needs to take in a noised image at a particular time step and return the predicted noise. Note that the predicted noise is a tensor that has the same size/resolution as the input image. So technically, the network takes in and outputs tensors of the same shape.[6] This task is called semantic segmentation. Given an image, assign every pixel (smallest atomic unit of an image) to a class that it belongs to.

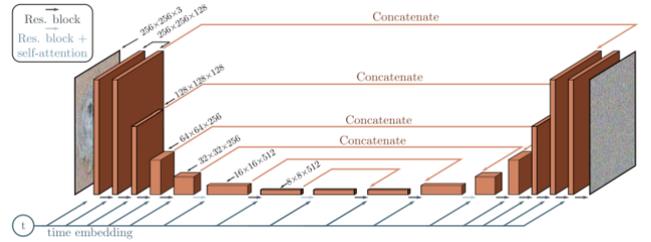

Figure 4: U net Architecture

The structure of this neural network is in the shape of a U. There's a reason for this symmetry. The first half of the "U" shrinks the input image and extracts meaningful information (called features) from it at each stage. The second half of the "U" reverses this process, expanding the intermediate results (called feature maps) to larger and larger sizes until the final output is the same size as the input image. To do so, it uses a layer known as a transposed convolution in addition to the pooling and activation layers. Our U-Net was designed to take a 256 by 256 image and return the same dimensions.

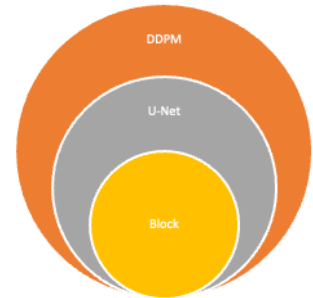

Figure 5: Overview of the overall architecture

Figure 5 represents the overall architecture in a bird's eye view for easy understanding.

### C. Sinusoidal Embedding

In order to share parameters across time (representing noise levels) in their neural network, the authors of the study use sinusoidal position embeddings. This idea is inspired by the Transformer model introduced in the work by Vaswani et al. (2017) [6]. By incorporating sinusoidal position embeddings, the neural network gains awareness of the specific time step (or noise level) it is processing for each image in a batch.

The SinusoidalPositionEmbeddings module takes a tensor of shape (batch_size, 1) as input, representing the noise levels of multiple noisy images in a batch. It transforms this tensor into a new one with shape (batch_size, dim), where 'dim' refers to the dimensionality of the position embeddings. This resulting tensor is then added to each residual block in the neural network, as we will discuss in more detail later on.

### D. System Setup

We used GPU-powered Google Colab for this project since it required a substantial amount of parallel processing. The dataset was hosted from a private google drive account, and the best models were saved in the same.

### E. Data Preprocessing

The images were having different dimensions, and the first step we performed was to reshape them to the desired 256 by 256-pixel dimension. Once that was done, we performed center cropping to remove noise that did not focus on the image. Thirdly, the cropped image was converted to a grey scale, thus giving only one channel to work with and reducing the complexity of the image. There will also be a lower information loss in this step since X-Ray images consist of black and white shades. Finally these images were made into a tensor and normalized in the range of [-1,1]. The reason that we normalize in the range of [-1,1] instead of [0,1] is the DDPM model's network predicts normally distributed noise throughout the denoising process.

### F. TRAINING PROCESS

The training objective of diffusion-based generative models amounts to "maximizing the log-likelihood of the sample generated (at the end of the reverse process) (x) belonging to the original data distribution." [7] or simply put, learn the reverse process - i.e. training $p_\theta(x_{t-1}|x_t)$. The methodology we have used is to first run the forward process to add noise from a Gaussian distribution to the selected batch of images, which will then give a tensor of (N,C,H,W) where N is the batch size followed by channels, height and width. Once we obtain the noisy images, we try to estimate the noise using the neural network with the U-Net architecture. Once we have the predicted noise and the actual added noise we use a simple MSE to adjust the parameters of the neural net accordingly. The network's gradients are set to 0 initially for each epoch to avoid gradients being stacked over the training process. Then, the backward function on the loss will compute the gradients and finally updates the parameters accordingly. In this project, we have trained the model for 50 epochs with a 0.001 learning rate and the best model was obtained at epoch 31 with a minimum epoch loss of 0.009. In each epoch, the performance is checked against the previous best loss and saved in a local directory as a model artifact.

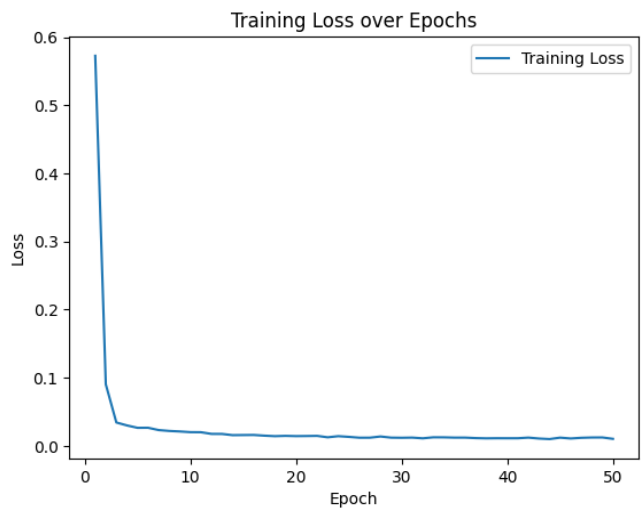

**Algorithm 1** Training
1: **repeat**
2: $\quad x_0 \sim q(x_0)$
3: $\quad t \sim \text{Uniform}(\{1, \ldots, T\})$
4: $\quad \epsilon \sim \mathcal{N}(0, I)$
5: $\quad$ Take gradient descent step on
$$\nabla_\theta \left\| \epsilon - \epsilon_\theta(\sqrt{\bar{\alpha}_t} x_0 + \sqrt{1 - \bar{\alpha}_t}\epsilon, t) \right\|^2$$
6: **until** converged

Figure 6: Algorithm of training steps

The process is summarized in Figure 6 [2] clearly to get an overview of the explained steps.

Figure 7: Training Loss

Figure 7 suggests that the model's rate of improvement has slowed from epoch 10 onwards. We can implement early stopping callback to avoid further training after around this points but in this project, we have not for research purposes.

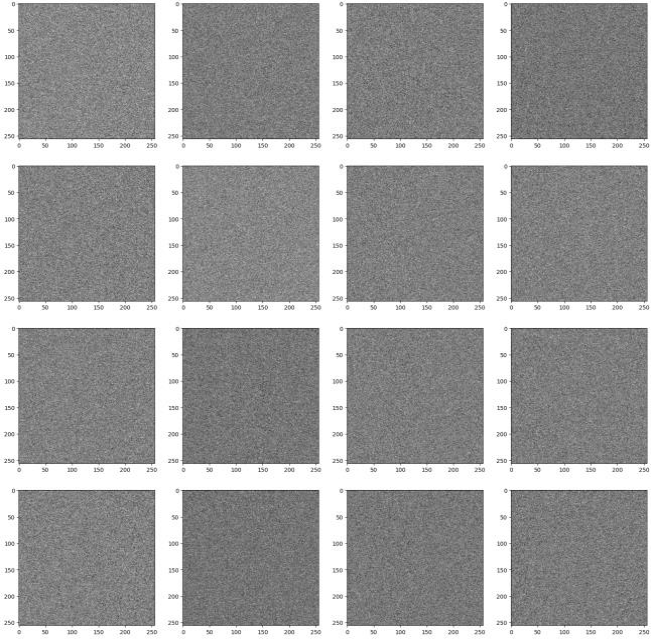

Figure 8: Completely noisy images before training

Figure 8 gives the images before training and you can see that we are starting with complete noise.

*G. EVALUATION*

Evaluating synthetically created images is not straightforward, and we can take different approaches to achieve this. Below are a few,
- Having visually examined by domain experts
- Using MSE between real images and synthetically generated
- Structural Similarity Index (SMI)
- Training an image classifier on real images with higher accuracy and checking the synthetic images on it

Out of those, we have performed MSE and SMI since they are more straightforward and we can obtain a score. We will discuss these two in detail,

MSE: In this method, the idea was to get a score based on the difference between the pixel values of the synthetically generated images and the real images. We tested this case for two more cases, where we tested the mse between synthetic pneumonia images and real images, which represents healthy cases without pneumonia, and the third was to check mse on two classes of the real images only. The idea was to get an understanding of how sparse the classes were.

SMI: The Structural Similarity Index (SSIM) metric extracts 3 key features from an image, Luminance, Contrast and Structure.

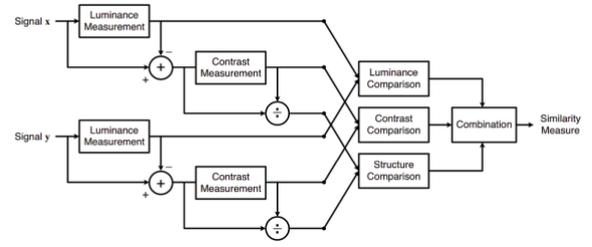

Figure 9: SMI

This system calculates the Structural Similarity Index between 2 given images which is a value between -1 and +1. A value of +1 indicates that the 2 given images are very similar or the same while a value of -1 indicates the 2 given images are very different. Often these values are adjusted to be in the range [0, 1], where the extremes hold the same meaning. [8] Below are the ways that the method calculates different aspects in detail,

Luminance: Luminance is measured by averaging over all the pixel values. Its denoted by μ (Mu) and the formula is given below,

$$\mu_x = \frac{1}{N}\sum_{i=1}^{N} x_i. \quad (8)$$

Contrast: It is measured by taking the standard deviation (square root of variance) of all the pixel values. It is denoted by σ (sigma) and represented by the formula below,

$$\sigma_x = \left(\frac{1}{N-1}\sum_{i=1}^{N}(x_i - \mu_x)^2\right)^{\frac{1}{2}}. \quad (9)$$

Structure: The structural comparison is done by using a consolidated formula (more on that later) but in essence, we divide the input signal with its standard deviation so that the result has unit standard deviation, which allows for a more robust comparison.

$$(\mathbf{x} - \hat{\mu}_x)/\sigma_x \quad (10)$$

The SMI score is given by combining all three aspects in the below formula,

$$\text{SSIM}(\mathbf{x}, \mathbf{y}) = [l(\mathbf{x}, \mathbf{y})]^\alpha \cdot [c(\mathbf{x}, \mathbf{y})]^\beta \cdot [s(\mathbf{x}, \mathbf{y})]^\gamma \quad (11)$$

where α > 0, β > 0, γ > 0 denote the relative importance of each of the metrics.

| Class 1 | Class 2 | MSE | SSMI |
|---|---|---|---|
| Real positive cases | Synthetic positive cases | 105.65 | 0.43 |
| Real negative cases | Synthetic positive cases | 104.94 | 0.28 |
| Real negative cases | Real positive cases | 194.63 | 0.26 |

Table 1: Evaluation matrix

## III. RESULTS

The trained model was able to generate some good results given that it was fed 256 by 256 low dimension images and only trained for 50 epochs. Figure 10 shows some of the synthetically generated pneumonia cases by the trained model.

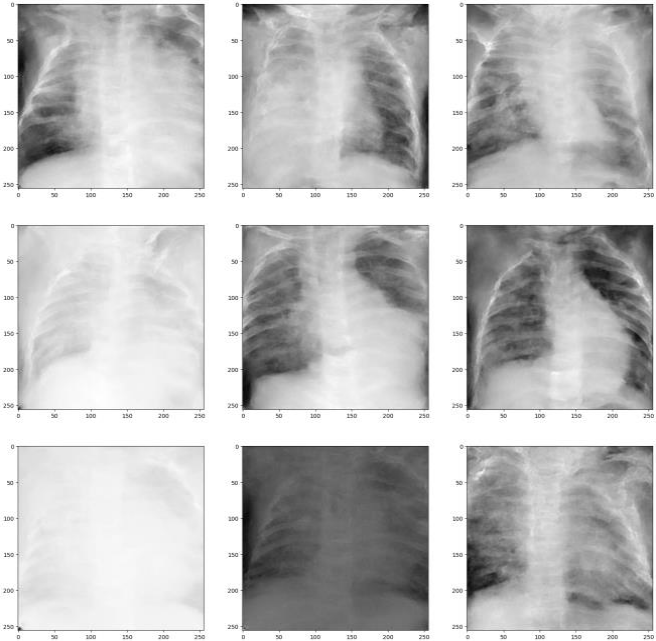

Figure 10: Model Generated images

As you can see, the model was able to generate ideal lung structures for most cases, but close inspection shows it has its limitations.

## IV. CONCLUSION

In this project, we addressed the challenge of limited positive medical image datasets, specifically focusing on pneumonia X-ray images. We proposed a novel approach using the Denoising Diffusion Probabilistic Model (DDPM) combined with a UNet architecture to synthesize realistic and diverse X-ray images of pneumonia. By leveraging advanced probabilistic models, our synthesizer successfully generated images that closely resembled real pneumonia cases. The results demonstrated the effectiveness of our approach, with the model generating high-quality images with low Mean Squared Error (MSE). These synthesized images exhibited distinct differences from non-pneumonia images, indicating the model's ability to capture key features of positive cases. The synthesized images hold promise in augmenting existing medical image datasets, thus aiding in training machine learning models for more accurate and efficient medical diagnoses.

Furthermore, our approach is not limited to pneumonia detection alone; it can be extended to various medical conditions, provided a sufficient dataset is available. The capability to produce high-quality synthetic images can significantly enhance the performance and accuracy of machine learning models in medical image analysis, leading to improved patient outcomes and more effective healthcare diagnostics. As with any innovative project, there are limitations to our approach. The model's performance is influenced by the quality and diversity of the training dataset. The scarcity of positive medical image datasets remains a challenge, and efforts to collect and curate larger and more diverse datasets should be pursued to further improve the synthesizer's performance..

## V. FUTURE WORK

The synthesis of realistic images has the potential to revolutionize medical image analysis, contributing to more accurate and efficient diagnoses in the healthcare industry. As future work, we will explore methods to improve the model's performance further and investigate its applications in generating synthetic images for other medical conditions beyond pneumonia. The combination of advanced generative models and medical imaging holds immense potential for transforming the field of healthcare and improving patient care.